\documentclass[aps,prd,amsfonts,amssymb,amsmath,nofootinbib,notitlepage,11pt,tightenlines,longbibliography]{revtex4-1}
\usepackage{breqn,esint,graphics,hyperref,array,xcolor,mathtools,longtable}

\newcommand{\tmcolor}[2]{{\color{#1}{#2}}}
\newcommand{\tmem}[1]{{\em #1\/}}

\newcommand{\tmstrong}[1]{\textbf{#1}}
\newcommand{\tmtextit}[1]{{\itshape{#1}}}

\makeatletter
\let\cat@comma@active\@empty
\makeatother

\begin{document}
\title{Finite N corrections to white hot string bits}
\author{Sourav Raha}
\email[Email: ]{souravraha@ufl.edu}
\affiliation{Institute for Fundamental Theory, Department of Physics,
University of Florida, Gainesville, Florida 32611, USA}
\date{\today}
\begin{abstract}
String bit systems exhibit a Hagedorn transition in the $N\to\infty$ 
limit. However, there is no phase transition when $N$ is finite (but 
still large). We calculate two-loop, finite $N$ corrections to the 
partition function in the low-temperature regime. The Haar measure in the singlet-restricted partition function contributes pieces to loop 
corrections that diverge as $\mathcal{O}(N)$ when summed over the 
mode numbers. We study how these divergent pieces cancel each other 
out when combined. The properly normalized two-loop corrections vanish 
as $\mathcal{O}(N^{-1})$ for all temperatures below the Hagedorn temperature. The coefficient of this $1/N$ dependence decreases with temperature and diverges at the Hagedorn pole.
\end{abstract}
\maketitle

\section{Introduction}
One can study a light-cone-quantized string as the continuum limit of a polymer
of point masses called string
bits {\cite{giles_lattice_1977,thorn_reformulating_1994}}. These bits move in
transverse space, enjoy nearest-neighbor interactions and transform adjointly
under a global $U (N)$ symmetry. It is possible to incorporate target space
supersymmetry {\cite{bergman_string_1995}} into this picture. The longitudinal
coordinate is recovered in the large $N${\cite{hooft_planar_1974}} limit and the
continuum limit of such a polymer. In fact, as an extreme form of holography,
one may recover all the coordinates (instead of simply the longitudinal one)
by postulating extra internal degrees of freedom (d.o.f) {\cite{thorn_space_2014}}. It
is instructive to study the behavior of such a system at finite
temperature {\cite{thorn_string_2015,raha_hagedorn_2017,curtright_color_2017}}.
Such systems exhibit a Hagedorn transition from a low-temperature phase that consists of closed chains to a high-temperature phase consisting of liberated bits. This bears similarities to Hagedorn transitions studied in
various other models: Hermitian matrix model {\cite{brezin_planar_1978}}, unitary matrix model {\cite{gross_possible_1980}} and $\mathcal{N}=4$ supersymmetric Yang-Mills (SYM) theory on $S^3$ {\cite{sundborg_hagedorn_2000,aharony_hagedorn/deconfinement_2004}}. For an improved calculation of Hagedorn temperature in $\mathcal{N}=4$ SYM theory at finite 't Hooft's coupling see {\cite{harmark_hagedorn_2018,harmark_hagedorn_2018-1}}.

In a recent paper {\cite{curtright_color_2017}} we have computed the
low-temperature partition function of the simplest stable
string bit system, up to leading order in $N$. In that simplified model, interactions between the string bits were switched off and instead, a singlet restriction was imposed on them. In the appropriate limits, this rudimentary system corresponds to the $T_0\to 0$ limit of a subcritical string in $1+1$ dimensions that has only one Grassmann world sheet field. We observed that the singlet
restriction can be studied as $1 / N$ perturbations in an effective scalar
field theory. The Hagedorn temperature of the system could then be understood
as the location of the pole of the ``bare propagator'' in this effective field
theory. At large but finite $N$ the system is not supposed to have a Hagedorn
transition (there are only a finite number of (d.o.f) at finite
$N$). This motivated us to do a partial resummation of the bare
propagator with quartic corrections to shift the Hagedorn pole off the real
temperature axis. Only at infinite temperature did we manage to compute finite
$N$ corrections to the partition function and discovered its link to an
enumeration problem of Eulerian digraphs with $N$ nodes. As a follow-up to our
paper, Beccaria used the technique developed in
{\cite{aharony_hagedorn/deconfinement_2004}} to calculate the density of
eigenvalues in the high-temperature phase up to leading order in
$N$ {\cite{beccaria_thermal_2017}}.

In this paper we shall present finite $N$ corrections to the following partition
function in the low-temperature regime:
\begin{widetext}
\begin{equation}
  Z = \dfrac{\left( \dfrac{1 + x}{1 - x} \right)^{N - 1} \left(\displaystyle\int^{\pi}_{-
  \pi} \displaystyle\prod_{k = 1}^N d \theta_k \right)  \left(
  \displaystyle\prod_{1 \leq i < j \leq N} 4 \sin^2 \left( \dfrac{\theta_i -
  \theta_j}{2} \right) \dfrac{|1+x e^{i (\theta_i-\theta_j)}|^{2f}}{|1-x e^{i (\theta_i-\theta_j)}|^{2b}} \right)}{\left( \displaystyle\int^{\pi}_{-
  \pi} \displaystyle\prod_{k = 1}^N d \theta_k \right)  \left(
  \displaystyle\prod_{1 \leq i < j \leq N} 4 \sin^2 \left( \dfrac{\theta_i -
  \theta_j}{2} \right) \right)}
\label{original}
\end{equation}
\end{widetext}
where $x = e^{- \beta \omega}$, $\theta_k$ represents the
$k$th angle in the $U(N)$ color space $\forall k \in \{ 1, N \}$, $b$ is the number
of distinct bosonic species and $f$ is the number of distinct fermionic
species in the system. $\beta$ denotes $\dfrac{1}{k_B T}$ and $\omega$ denotes the mass of a string bit. Our results will hold for $0 \leqslant x < \dfrac{1}{b
+ f}$. The connected vacuum diagrams are then represented by
\begin{dmath}
  \log (Z) = \log \int^{\pi}_{- \pi} \prod^N_{k = 1} d \theta_k \exp [L (x ;
  \{ \theta \})] - \log \int^{\pi}_{- \pi} \prod^N_{k = 1} d \theta_k \exp [L
  (0 ; \{ \theta \})]
\end{dmath}
\label{second}where
\begin{equation}
  L (x ; \{ \theta \}) = (N - 1) \log \left( \dfrac{1 + x}{1 - x} \right) +
  \dfrac{1}{2} \sum_{i \neq j} \mathcal{L} (x ; \theta_i - \theta_j)
\end{equation}
with
\begin{equation}
  \mathcal{L} (x ; \theta) = \log (1 - e^{i \theta}) + f \log (1 + x e^{i
  \theta}) - b \log (1 - x e^{i \theta}) + c.c.
\label{3pieces}
\end{equation}
containing pieces from the group measure, fermionic bits and
bosonic bits, respectively. In the low-temperature phase, $L$ is maximized by
a uniform distribution, $\theta_0$, of $\{ \theta \}$. One can take a
nondecreasing function of the indices,
\begin{align}
  {\theta_{0_{k}}} &= 2 \pi \dfrac{k}{N} & k &\in \{ 1, \ldots, N \}
\end{align}
and expand this effective Lagrangian about this uniform distribution. Then
using perturbation theory for scalar field
\begin{widetext}
\begin{dmath}
  \log \left( \int e^L \right)  =  L_0 + \frac{1}{2} \log \left( \det \left[
  \dfrac{2 \pi}{- L_2} \right] \right)
  + \left. \log \left[ \exp \left\{ \dfrac{L_3}{3!} \left(
  \dfrac{\delta}{\delta J} \right)^3 + \dfrac{L_4}{4!} \left(
  \dfrac{\delta}{\delta J} \right)^4 + \cdots \right\} \exp \left\{
  \dfrac{J^2}{2 (- L_2)} \right\} \right] \right|_{J = 0}\\
  \approx  L_0 + \frac{1}{2} \log \left( \det \left[ \dfrac{2 \pi}{- L_2}
  \right] \right)
  - \dfrac{1}{12}  \sum_{m,n} \dfrac{V_{m, n, - m - n} V_{- m, - n, m + n}}{V_{m, -
  m} V_{n, - n} V_{- m - n, m + n}} + \frac{1}{8}  \sum_{m,n} \dfrac{V_{m, - m, n, -
  n}}{V_{m, - m} V_{n, - n}} + \cdots
\label{vacc}
\end{dmath}
\end{widetext}
where $L_p \equiv L_{k_1, \ldots, k_p} = \dfrac{\delta^p L
[\theta_0]}{\delta \theta_{k_1} \cdots \delta \theta_{k_p}}$ are the coupling
constants in ``position space'' and
\begin{equation}
  V_{n_1, \ldots, n_p} = \dfrac{1}{N^{p / 2}} \sum^N_{k_1, \cdots, k_p = 1}
  L_{k_1, \ldots, k_p} e^{2 \pi i (n_1 k_1 + \cdots + n_p k_p) / N}
\end{equation}
are the coupling constants in ``Fourier space.'' $L_2 [\theta_0]$ turns out to
be a circulant matrix in the ``position indices,'' i.e. $L_{m, n} [\theta_0] =
F (| m - n |)$, and hence can be naturally diagonalized via the Fourier
transform {\cite{curtright_color_2017}}.

\section{Calculation of vertices for finite $N$}
In {\cite{curtright_color_2017}}, the $p$th Fourier vertex is given by
\begin{equation}
  V_{n_1 \cdots n_p} = \frac{\delta_{N|n_1 + \cdots + n_p}}{2 N^{1 - p / 2}}
  \sum^{N - 1}_{\alpha = 1} \frac{d^p \mathcal{L} \left( 2 \pi
  \dfrac{\alpha}{N} \right)}{d \theta^p}  (e^{2 \pi i \alpha n_1 / N} - 1)
  \cdots (e^{2 \pi i \alpha n_p / N} - 1)
\end{equation}
where $n_l \in \mathbb{Z} \forall l \in \{ 1, p \}$ represents the Fourier
mode numbers, and the delta symbol is $1$ whenever $N$ is a factor of $n_1 +
\cdots + n_p$ and $0$ otherwise. $\mathcal{L}$ is a function of differences in
$\theta$'s, hence its derivative with respect to a single $\theta_k$ yields
differences in Kronecker deltas:
\begin{equation}
  \dfrac{d \mathcal{L}}{d \theta_k} = \sum_{i \neq j} (\delta_{i k} -
  \delta_{j k})  \mathcal{L}' (\theta_k)
\end{equation}
Upon a Fourier transform these differences in Kronecker deltas yield products
of differences between powers of roots of unity:$(e^{2 \pi i \alpha n_1 / N} -
e^{2 \pi i \beta n_1 / N}) \cdots (e^{2 \pi i \alpha n_p / N} - e^{2 \pi i
\beta n_p / N})$. Following this, in \cite{curtright_color_2017} we approximated the sum over $\alpha$ by an
integral. In this paper, we shall perform the exact summation.

But first, let us try to evaluate the following expression
\begin{dmath}
  \mathcal{B} (\{ n \} ; t)  =  \sum_{\alpha} \left. (e^{i n_1
  \theta_{\alpha}} - 1) \cdots (e^{i n_p \theta_{\alpha}} - 1) \left(
  \dfrac{d}{d \theta_{\alpha}} \right)^p \log (1 - e^{t + i \theta_{\alpha}})
  \right|_{\theta_k=2\pi\tfrac{k}{N}}\\
   =  \left( i \dfrac{d}{d t} \right)^p \sum_{\alpha} (e^{2 \pi i n_1
  \alpha / N} - 1) \cdots (e^{2 \pi i n_p \alpha / N} - 1) \log (1 - e^{t + 2
  \pi i \alpha / N})
\end{dmath}
where $\{ n \} \equiv \{ n_1, \cdots, n_p \}$ and in the first line we are
evaluating the entire summand at uniform distribution, $\theta_0$. The
derivative with respect to any $\theta_{\alpha}$ can be replaced by $i \dfrac{d}{d t}$ .
This enables one to pull the derivative operator outside the sum. This leaves
the sum to be independent of the order, $p$, of the vertex. One can generate
any vertex by repeatedly applying $i \dfrac{d}{d t}$ on this universal sum.
Expanding the logarithm on rhs we get
\begin{dmath}
  \mathcal{B} (\{ n \} ; t)  =  - \left( i \dfrac{d}{d t} \right)^p \sum_{m
  = 1}^{\infty} \dfrac{e^{m t}}{m}  \sum_{\alpha = 1}^{N - 1} e^{2 \pi i m
  \alpha / N}  (e^{2 \pi i n_1 \alpha / N} - 1) \cdots (e^{2 \pi i n_p \alpha
  / N} - 1) \\
   =  - i^p \left( \dfrac{d}{d t} \right)^{p - 1} \sum_{m = 1}^{\infty}
  e^{m t} \sum_{\alpha = 1}^{N - 1} \sum_{s \in \{ n \}} C_{y (s)} e^{2 \pi i
  (m + \widetilde{y (s)}) \alpha / N} 
\end{dmath}
where,
\begin{equation}
  (e^{2 \pi i n_1 \alpha / N} - 1) \cdots (e^{2 \pi i n_p \alpha / N} - 1) =
  \sum_{s \in \{ n \}} C_{y (s)} e^{2 \pi i \widetilde{y (s)} \alpha / N}
\end{equation}
with $y (s)$ denoting the sum total of the elements in a {\tmem{subset}} $s$
of $\{ n \}$. For example, $y (s)$ could represent $(n_1 + n_5)$
, $(n_2 + n_3 + n_{p -
1})$, etc. $C_{y (s)} \in \{ - 1, + 1 \}$ is the coefficient corresponding to
a particular $s$ and $\widetilde{y (s)} \equiv y (s) \mod N$. The sum
over $s$ represents a sum over all possible ways of obtaining subsets from $\{
n \}$. Finally, we have
\begin{equation}
  \mathcal{B} (\{ n \} ; t) = - i^p \left( \dfrac{d}{d t} \right)^{p - 1}
  \sum_{s \in \{ n \}} C_{y (s)}  \left\{ \dfrac{N e^{t (N - \widetilde{y
  (s)})}}{1 - e^{t N}} - \dfrac{e^t}{1 - e^t} \right\}
\end{equation}
The presence of the mod function (represented by $\widetilde{\  }$ ) tells one
that $\mathcal{B} (\{ n \} ; t)$ is periodic in each value of $n$. 
Now one can express $V$'s in a very compact form in terms of these
$\mathcal{B}$'s:
\begin{dmath}
  V_{n_1, \cdots, n_p} = \frac{\delta_{N|n_1 + \cdots + n_p}}{2 N^{1 - p / 2}}
  \left\{ \displaystyle\lim_{t \rightarrow 0^-}  \mathcal{B} (\{ n \} ; t) +
  \displaystyle\lim_{t \rightarrow 0^-} \mathcal{B} (\{ - n \} ; t)^{\ast}
   \\
  + f \mathcal{B} \left( \{ n \} ; - \beta \omega + i \pi
  \right) + f \mathcal{B} (\{ - n \} ; - \beta \omega + i \pi)^{\ast}
  \\
   - b \mathcal{B} (\{ n \} ; - \beta \omega) - b \mathcal{B}
  (\{ - n \} ; - \beta \omega)^{\ast} \right\} \label{vertex}
\end{dmath}
where $\{ - n \} \equiv \{ - n_1, \cdots, - n_p \}$ and ``$^{\ast}$'' denotes
complex conjugation.

This is the formula with which we may compute any diagram at finite $N$. In a
parallel to Eq. [\ref{3pieces}], one can verify that the terms in the first
row account for the contribution from the group measure, the second row
accounts for the adjoint fermions and the third row for the adjoint bosons.
The (bare) inverse propagator then is\footnote{From here onward, we shall
express everything in terms of $x = e^{- \beta \omega}$ instead of $\beta$. }
given by
\begin{dmath}
  V_{n, N - n} = \left\{ N^2  \left( b \dfrac{ (1 - x^n)^2 x^{N - n}}{(1 -
  x^N)^2} - f \dfrac{(1 - (- x)^n)^2 (- x)^{N - n}}{(1 - (- x)^N)^2} \right)  
    - N n \left( b \dfrac{x^n - x^{N - n}}{1 -
  x^N} - f \dfrac{ (- x)^n - (- x)^{N - n}}{1 - (- x)^N} + 1 \right) + n^2
  \right\}
\label{2vertex}
\end{dmath}
where $N > n > 0$. With some inspection one can confirm that
the rhs of this equation is symmetric under the interchange of $n
\leftrightarrow N - n$. With finite $n$, as $N \rightarrow \infty$,
\begin{equation}
  V_{n, N - n} \approx - N n \{ 1 - b x^n + f (- x)^n \} = - N n I_n
\label{Nvertex2}
\end{equation}
which reproduces the (leading order in $N$) result obtained in 
{\cite{curtright_color_2017}}, with $I_n$ denoting the temperature dependent
factor.

At $x = 0$ (zero temperature) Eq. [\ref{2vertex}] gives $V_{n, N - n} = n (n
- N)$. Physically, this signifies the (bare) inverse propagator when the
integrand in
\begin{equation}
  \mathcal{I} = \int_{- \pi}^{\pi} \prod_{k = 1}^N d \theta_k  \prod_{i < j} 4
  \sin^2 \left( \dfrac{\theta_i - \theta_j}{2} \right)
\label{haar}
\end{equation}
is expanded about the $\theta_0$ (the uniform distribution, also
the global maximum). $\mathcal{I}$ is the same as the normalization on the rhs
of Eq. [\ref{original}]. Using the method of steepest descent, one may calculate
this integral to be
\begin{dmath}
  \mathcal{I} = N! \exp \left\{ N \log (N) + \log (2 \pi) + \dfrac{N - 1}{2}
  \log (2 \pi) - \log ((N - 1) !) + \Lambda \right\}
\end{dmath}
where the prefactor is due to the permutation symmetry of the integrand under
the exchange of $i, j$ indices. The first term in the parentheses indicates
the value of the integrand at uniform distribution while the next three terms
come from Gaussian fluctuations about the uniform distribution. However, $\mathcal{I}$ can be computed exactly and is known to be equal to $N! (2 \pi)^N$. We can use
the known answer to estimate the asymptotics of the higher order (i.e. beyond
Gaussian) corrections,
\begin{dmath}
  \Lambda  = - N \log (N) + \dfrac{N - 1}{2} \log (2 \pi) + \log ((N - 1)
  !)   =  - \left( 1 - \dfrac{\log (2 \pi)}{2} \right) N - \dfrac{1}{2} \log
  (N) + \cdots
\end{dmath}
This shows that Gaussian fluctuations are not enough to approximate
$\mathcal{I}$ as $N \rightarrow \infty$. We can numerically show that one
needs to take into account \tmtextit{at least} the two-loop corrections in
order to obtain the correct (large $N$) limit.

\begin{figure}[h]
  \resizebox{\columnwidth}{!}{\includegraphics{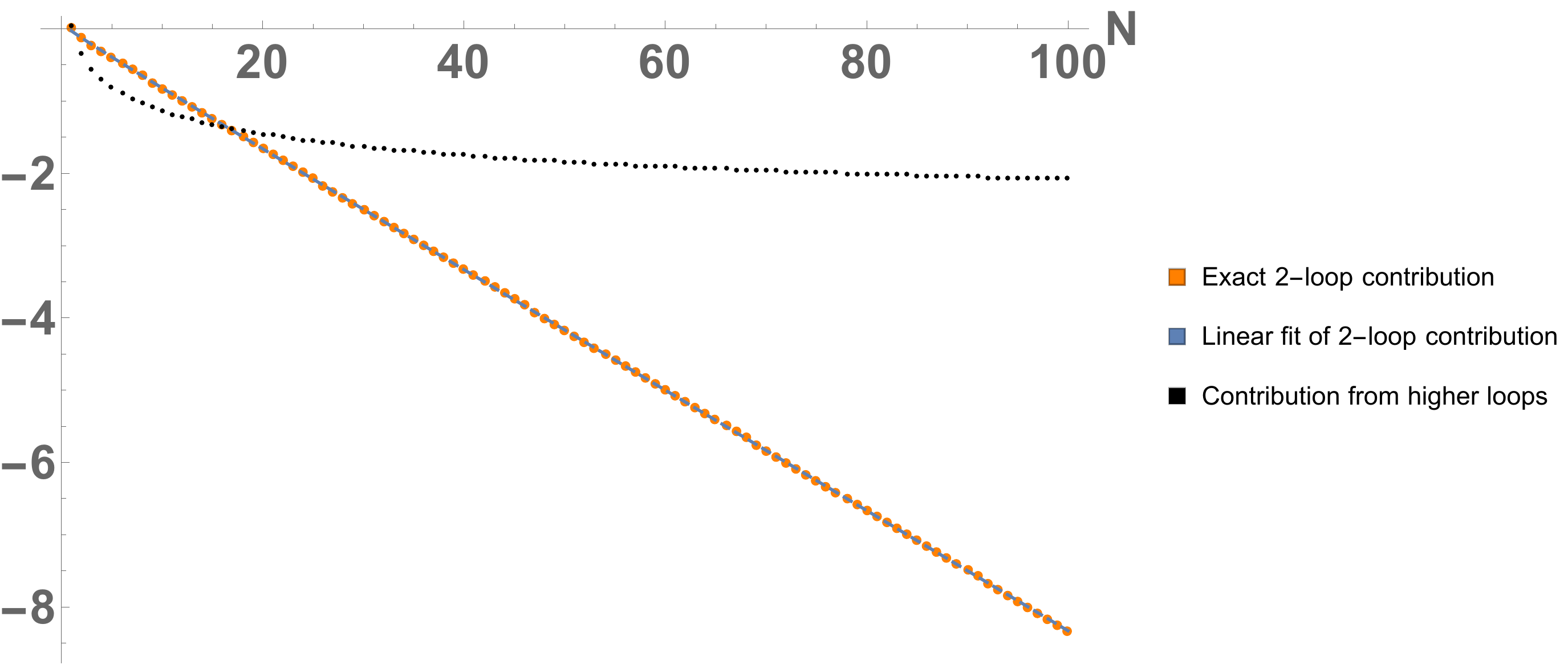}}
  \caption{Contribution of loop diagrams to the log of the integral of the
  Haar measure, shown as a function of $N$. Two-loop corrections seem to
  account for the linear divergence of beyond-Gaussian contributions. The
  slope of the linear fit to the two-loop contribution is $- 0.08297 \pm
  0.00003$.}
\label{measure_diag}
\end{figure}
Fig. [\ref{measure_diag}] tells us that the two-loop diagrams contribute up to $\mathcal{O}
(N)$ to the integral in Eq. [\ref{haar}]. The higher-loop diagrams \tmtextit{seem} to
contribute only up to subleading $\mathcal{O} (\log (N))$ corrections.
However, the numerical error in estimating the slope is too small. That is,
even though the diagram may suggest otherwise, higher-loop diagrams do have a
minuscule contribution to the linear divergence. Coming back to Eq. [\ref{original}], one
cannot rule out a similar thing from happening to $\log (Z)$ either. This
integral, too, may end up with beyond-Gaussian corrections that consist of
pieces that diverge for large $N$. This would, in turn, challenge our
assumption \cite{raha_hagedorn_2017,curtright_color_2017} that stopping at Gaussian fluctuations is enough to accurately
compute large $N$ behavior of $\log (Z)$. At this point one may think of
normalization in the definition of Eq. [\ref{original}] and expect it to remove these
potential divergences at loop order. However, one cannot be sure of such a
cancellation {\tmem{a priori}}. One can analyze e.g., the double-cubic term on
the rhs of Eq. [\ref{vacc}] to see why. If each vertex had a divergent component of a
certain order, a rational function of these vertices (which is what a Feynman
diagram is) would give rise to new divergences. One can expect normalization
to remove the leading divergence in such a function. But divergences of
next-to-leading order, arising from the ``cross terms'' in the Feynman
diagram, could still be left unaltered\footnote{This
will become clearer in the upcoming section where the vertices are explicitly
mentioned.}. To the best of our knowledge, one
cannot guarantee the cancellation of these subleading divergences in an
arbitrary Lagrangian. The fate of these divergences has to be found out by
explicitly computing the 2-loop corrections. Thus we are motivated to
calculate corrections to $\log (Z)$ up to two-loop order in the following
section.

\section{Two-loop Corrections}
\begin{figure}[h]
  \resizebox{\columnwidth}{!}{
    \centering
     \begin{tabular}{|c|c|c|}
       \hline
        \resizebox{.2\columnwidth}{!}{\includegraphics{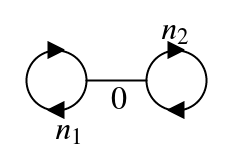}} &
        \resizebox{.11\columnwidth}{!}{\includegraphics{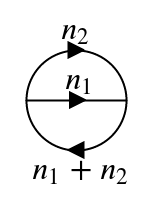}} &
        \resizebox{.14\columnwidth}{!}{\includegraphics{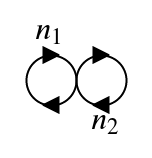}}\\
        \hline
      \end{tabular}}
  \caption{The possible two-loop corrections to the Gaussian result. In the
  system under consideration, the dumbbell diagram does not contribute.}
\label{unused}
\end{figure}

Like the (bare) inverse propagator in the previous section, one can compute
any (bare) vertex for finite values of $N$. For two-loop corrections one needs
expressions for the cubic and the quartic vertices only. There is no
contribution of the two-loop ``dumbbell'' diagram{\tmstrong{}}; as we see in
{\cite{curtright_color_2017}} that $V_{n, - n, 0}$ vanishes for our system.
The only contributions are from the ``theta'' (double-cubic) and the
``infinity'' (quartic) diagrams (see Fig. [\ref{unused}]).

\subsection{Cubic contribution}
The cubic vertex, using Eq. [\ref{vertex}], is represented by
\begin{widetext}
\begin{dmath}
  V_{n_1, n_2, - n_1 - n_2} = \left\{ b \sqrt{N} i \left( N^2  (1 + x^N) 
  \frac{\left( x^{N - (\widetilde{- n_1})} - x^{N - \widetilde{n_1}}
  \right)}{(1 - x^N)^3} - 2 N \frac{(\widetilde{- n_1}) x^{N - (\widetilde{-
  n_1})} - \widetilde{n_1} x^{N - \widetilde{n_1}}}{(1 - x^N)^2}
    + \frac{(\widetilde{- n_1})^2 x^{N - (\widetilde{- n_1})} -
  (\widetilde{n_1})^2 x^{N - \widetilde{n_1}}}{1 - x^N} \right) \\
  + \tmcolor{blue}{\{ b \rightarrow - f, x \rightarrow - x
  \}}   \\
  - \frac{i}{6 \sqrt{N}} \left( N^2  \text{} \left( \left(
  \widetilde{- n_1} \right) - \widetilde{n_1} \right) - 3 N ((\widetilde{-
  n_1})^2 - (\widetilde{n_1})^2) + 2 ((\widetilde{- n_1})^3 -
  (\widetilde{n_1})^3) \right)   \\
    + \tmcolor{blue}{\{ n_1 \rightarrow n_2 \}} +
  \tmcolor{blue}{\{ n_1 \rightarrow - n_1 - n_2 \}} \right\}
\end{dmath}
\end{widetext}
Here the third line on the rhs indicates that the fermionic contribution is
obtained by making corresponding substitutions in {\tmem{all}} the lines above it.
Similarly, the fifth line indicates that one obtains two more copies by making
the suggested substitutions in {\tmem{all}} the lines preceding it. Besides
the explicit symmetry under the permutations of the indices, one can see that
the rhs is an odd function of the indices. In other words
\begin{equation}
  V_{- n_1, - n_2, n_1 + n_2} = - V_{n_1, n_2, - n_1 - n_2} = V_{n_1, n_2, -
  n_1 - n_2}^{\ast}
\end{equation}
The contribution of the theta diagram is given by
\begin{equation}
  \Lambda_3 = - \dfrac{1}{12}  \sum_{n_1, n_2 = 1}^N \left( \dfrac{V_{n_1,
  n_2, - n_1 - n_2} V_{- n_1, - n_2, n_1 + n_2}}{V_{n_1, - n_1} V_{n_2, - n_2}
  V_{- n_1 - n_2, n_1 + n_2}} - \tmcolor{blue}{\{ x \rightarrow 0 \}} \right)
\label{sum_cub}
\end{equation}
where the subtracted part ensures proper, order-by-order
normalization of the diagram. An order-by-order normalization implies that one
is expanding the denominator on the rhs of Eq. [\ref{original}) about the
uniform distribution as well. This ensures the correct normalization of $Z$,
i.e. at $x = 0$, $Z$ will equal $1$ up to any order in perturbation. An
expansion of the summand of $\Lambda_3$ is neither compact nor illuminating.
It is far more useful to see it graphically

\begin{figure}[h]
  \resizebox{\columnwidth}{!}{\includegraphics{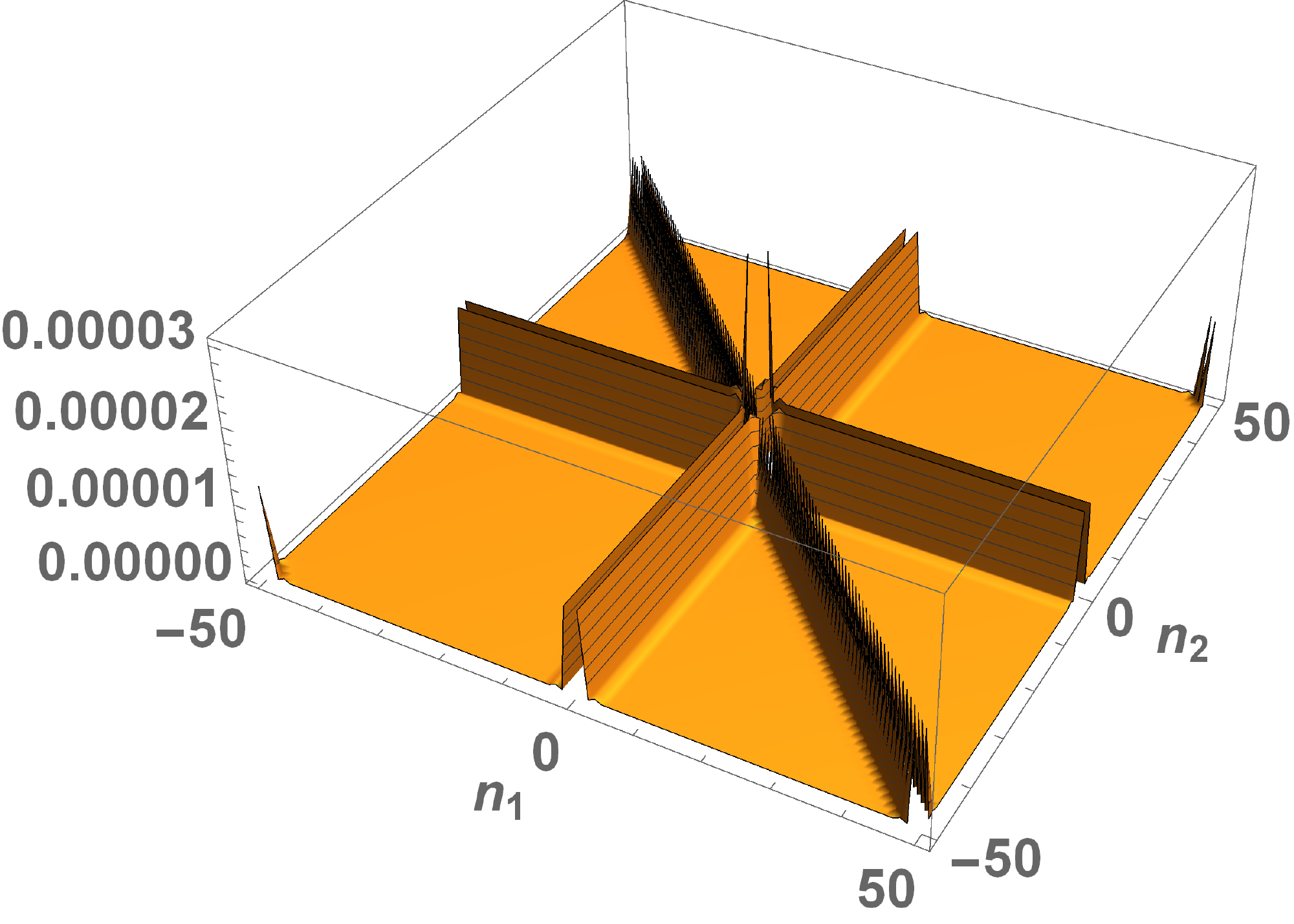}}
  \caption{Dependence on loop momenta $n_1$ \& $n_2$ of the summand of the
  cubic two-loop diagram for $N = 100$, $x = 0.125$, $b = f = 1$. The
  discontinuities represent regions where the propagators vanish, i.e. the
  zero modes.}
\label{cubic_diag}
\end{figure}

\begin{figure}[h]
  \resizebox{\columnwidth}{!}{\includegraphics{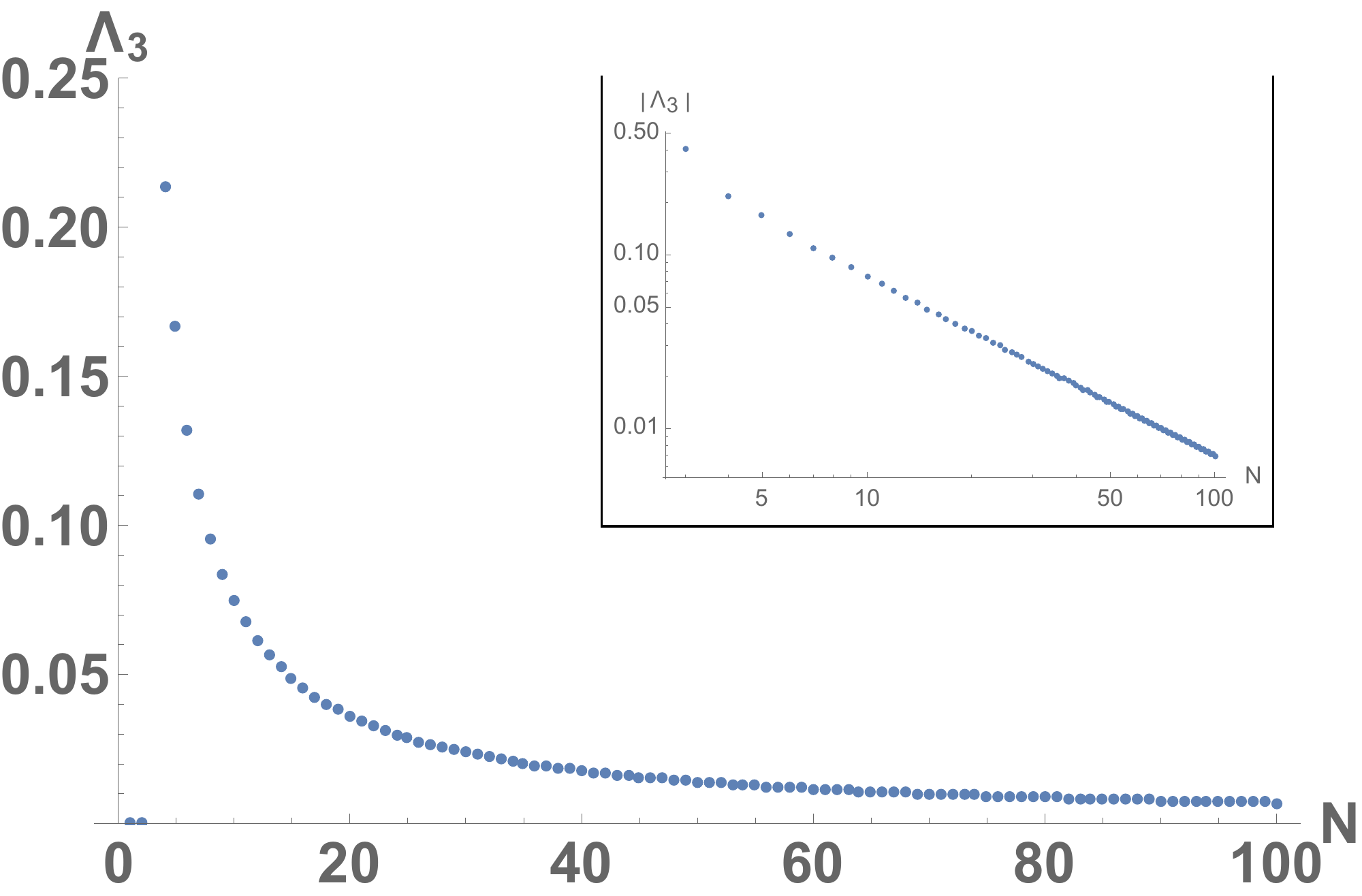}}
  \caption{Dependence on $N$ of the two-loop double-cubic corrections to the log
  of the partition function at $x = 0.125$ for $b = f = 1$. The inset displays
  the log-log scaled version of the main plot and has a slope of $- 1.057 \pm
  0.005$. This indicates that $\Lambda_3 \sim \mathcal{O} (N^{- 1})$.}
\label{cubvsN}
\end{figure}

Fig. [\ref{cubic_diag}] indicates that the double summation over loop momenta
yields an extra factor of $N$. One can deduce this from the presence of
ridge lines along $n_1, n_2 \approx \pm 1$. The ridges have constant
(nonzero) height and width, which leads to an extra factor of $N$ upon
summation. While it would be desirable to obtain a closed-form expression for
$\Lambda_3$, one can nonetheless employ numerics to study its behavior. Fig.[\ref{cubvsN}] is what one gets as one proceeds to actually plot the contribution of this diagram (after summing over
all the modes). This plot suggests that the {\tmem{summand}} of $\Lambda_3$ is $\mathcal{O}
(N^{- 2})$. That way, $\Lambda_3$ can vanish as $\dfrac{1}{N}$, despite an extra
factor of $N$ produced {\tmem{after}} the sum over modes.

\subsection{Quartic contribution}
Similarly, using Eq. [\ref{vertex}] for the quartic vertex one obtains:
\begin{widetext}
\begin{dmath}
 V_{n_1, n_2, - n_1, - n_2} = \left\{ 2 b \left( - N^3  (1 + 4 x^N + x^{2
   N})  \dfrac{x^{N - (\widetilde{- n_1})} + x^{N - \widetilde{n_1}}}{(1 -
   x^N)^4}
   + 3 N^2 (1 + x^N)  \frac{ (\widetilde{- n_1}) x^{N -
   (\widetilde{- n_1})} + \widetilde{n_1} x^{N - \widetilde{n_1}}}{(1 -
   x^N)^3} 
    - 3 N \frac{(\widetilde{- n_1})^2 x^{N - (\widetilde{- n_1})} +
   (\widetilde{n_1})^2 x^{N - \widetilde{n_1}}}{(1 - x^N)^2}  
        + \frac{(\widetilde{- n_1})^3 x^{N - (\widetilde{- n_1})}
   + (\widetilde{n_1})^3 x^{N - \widetilde{n_1}}}{1 - x^N} \right) \\
     + \tmcolor{blue}{\{ b \rightarrow - f, x \rightarrow - x
   \}}   \\
    - N \dfrac{ (\widetilde{- n_1})^2 +
   (\widetilde{n_1})^2}{2} + (\widetilde{- n_1})^3 + (\widetilde{n_1})^3 -
   \dfrac{\text{} (\widetilde{- n_1})^4 + (\widetilde{n_1})^4}{2 N}
     \\
     + \tmcolor{blue}{\{ n_1 \rightarrow n_2 \}} - \dfrac{1}{2}
   \tmcolor{blue}{\{ n_1 \rightarrow n_1 + n_2 \}} - \frac{1}{2}
   \tmcolor{blue}{\{ n_1 \rightarrow n_1 - n_2 \}}   \\
     + 4 N^3 \left( \dfrac{b x^N (1 + 4 x^N + x^{2 N})}{(1 -
   x^N)^4} - \dfrac{f (- x)^N (1 - 4 (- x)^N + x^{2 N})}{(1 + (- x)^N)^4}
   \right) \right\}
\end{dmath}
\end{widetext}
Just like the cubic vertex, one obtains various parts by making indicated
substitutions in {\tmem{all}} the lines preceding said substitution. Here,
because of the symmetry under the permutations of the indices, one can see
that the rhs is an even function in the indices. The contribution from quartic
correction to $\log (Z)$ is given by a double sum over $n_1$ and $n_2$. It is
given by
\begin{equation}
  \Lambda_4 = \dfrac{1}{8}  \sum_{n_1, n_2 = 1}^N \left( \dfrac{V_{n_1, - n_1,
  n_2, - n_2}}{V_{n_1, - n_1} V_{n_2, - n_2}} - \tmcolor{blue}{\{ x
  \rightarrow 0 \}} \right)
\label{sum_qrt}
\end{equation}
again, the expression being appropriately normalized.

\begin{figure}[h]
  \resizebox{\columnwidth}{!}{\includegraphics{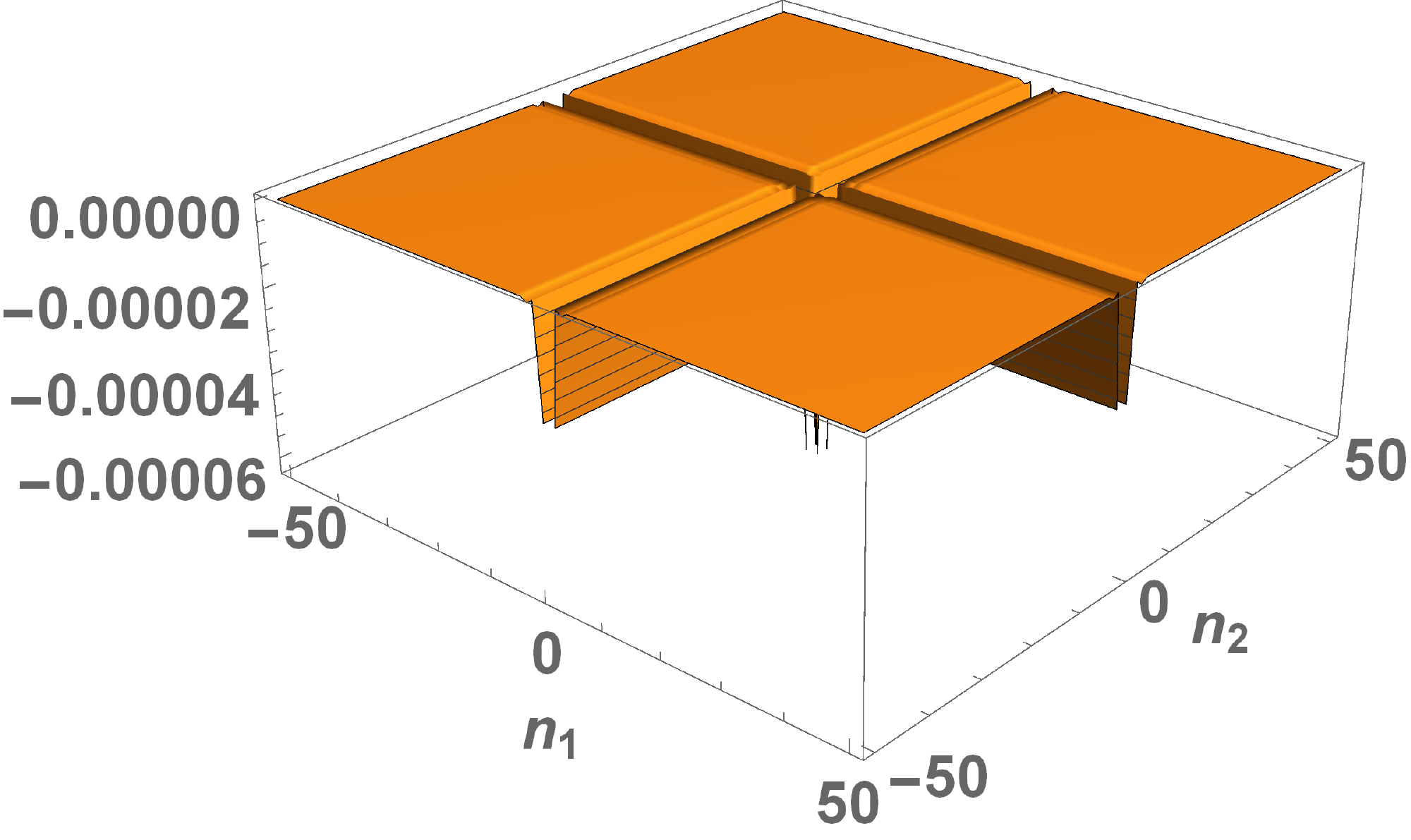}}
  \caption{Dependence on loop momenta $n_1$ and $n_2$ of the summand of the
  quartic two-loop diagram for $N = 100$, $x = 0.125$, $b = f = 1$. The
  discontinuities represent regions where the propagators vanish.}
\label{quartic_diag}
\end{figure}
\begin{figure}[h]
  \resizebox{\columnwidth}{!}{\includegraphics{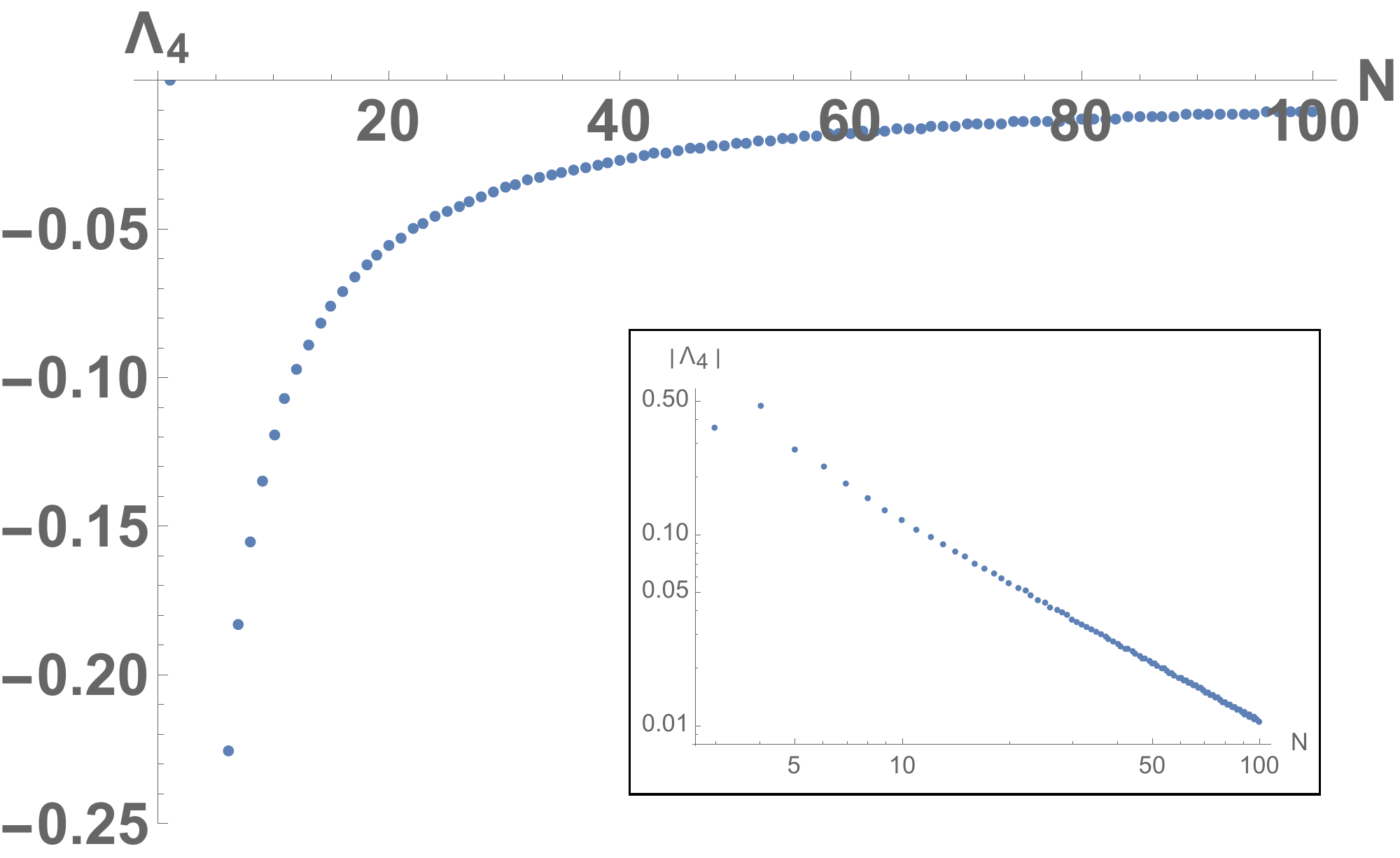}}{\tmem{}}
  \caption{Dependence on $N$ of the two-loop quartic corrections to the
  log of the partition function at $x = 0.125$ for $b = f = 1$. The inset
  displays a log-log scaled version of the main plot and has a slope of $-
  1.071{\pm}0.006$. This indicates that $\Lambda_4 \sim \mathcal{O} (N^{-
  1})$.}
\label{qrtvsN}
\end{figure}

In a parallel to the previous subsection, one may examine Fig.
[\ref{quartic_diag}] and see a factor of $N$ in it. Again, the primary
contribution to the sum comes from regions that have small mode number, i.e.
$| n_1 | = | n_2 | \approx 1$. And just like the cubic case, Fig. [\ref{qrtvsN}] reveals that the summand in $\Lambda_4$ too goes as $\mathcal{O} (N^{-
2})$.

In the next section, we shall discuss the reasons for this large $N$
dependence. We shall identify pieces in the summand of each diagram that
diverge on their own. And we shall try to see which pieces cancel each other
out when combined.

\section{Discussions}
\begin{figure}[h]
  \resizebox{\columnwidth}{!}{\includegraphics{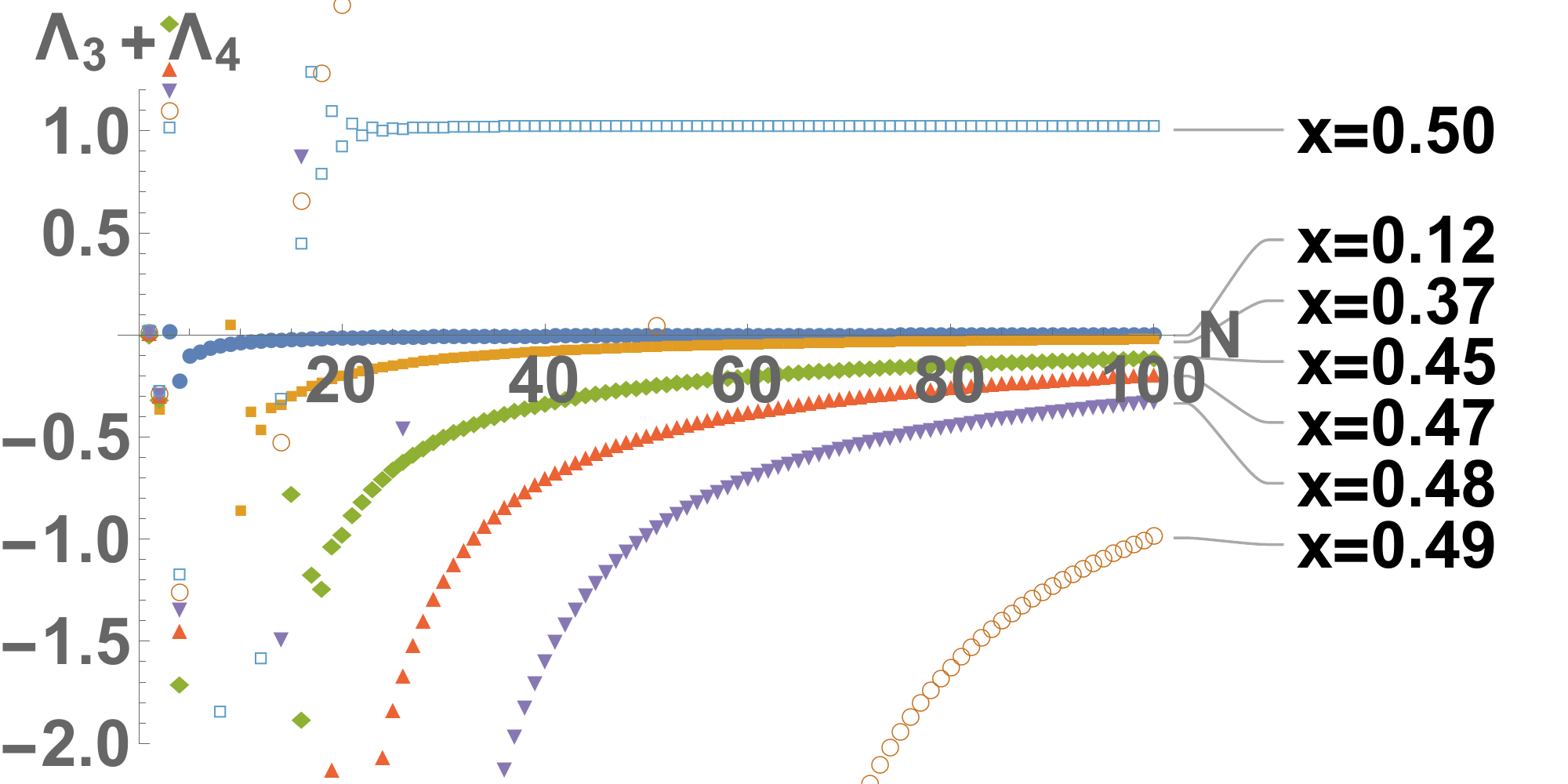}}
  \caption{Full two-loop corrections to the log of the partition function at
  different values of $x$ for $b = f = 1$ as functions of $N$. For $x < \dfrac{1}{b+f}$,
  the two-loop contribution is negative and vanishes as $\mathcal{O} (N^{-
  1})$ for large $N$.}
\label{total}
\end{figure}

In Fig. [\ref{total}] we have plotted full two-loop corrections for different values of
$x$ (but each below $x_H$). In the low-temperature phase,
one-loop corrections are already included in the \tmtextit{infinite} $N$
result, and the two-loop corrections are the leading \tmtextit{finite} $N$
corrections. One can immediately notice that the latter are negative
for large enough $N$. In [\cite{raha_hagedorn_2017}] we have presented exact $\log (Z)$ as analytic
functions of temperature, for a few small values of $N$. At least for those
cases, $\log (Z)$ seems to increase with $N$. Since $\log (Z)$ is not known as
an analytic function of $N$ it is not obvious whether this trend continues for
higher values of $N$. The negative sign of the (leading) finite $N$
corrections is, however, consistent with such a trend. There is obviously no
phase transition in a system with finite degrees of freedom. One can see this
in [\cite{raha_hagedorn_2017}] where exact partition functions for small $N$ have no divergence except
at infinite temperature. Because of the negative sign, the two-loop
corrections assist in pushing the Hagedorn pole to a higher temperature.
Including finite $N$ corrections to all loop orders would eventually push the
corresponding Hagedorn temperature to infinity.

\begin{figure}[h]
  \resizebox{\columnwidth}{!}{\includegraphics{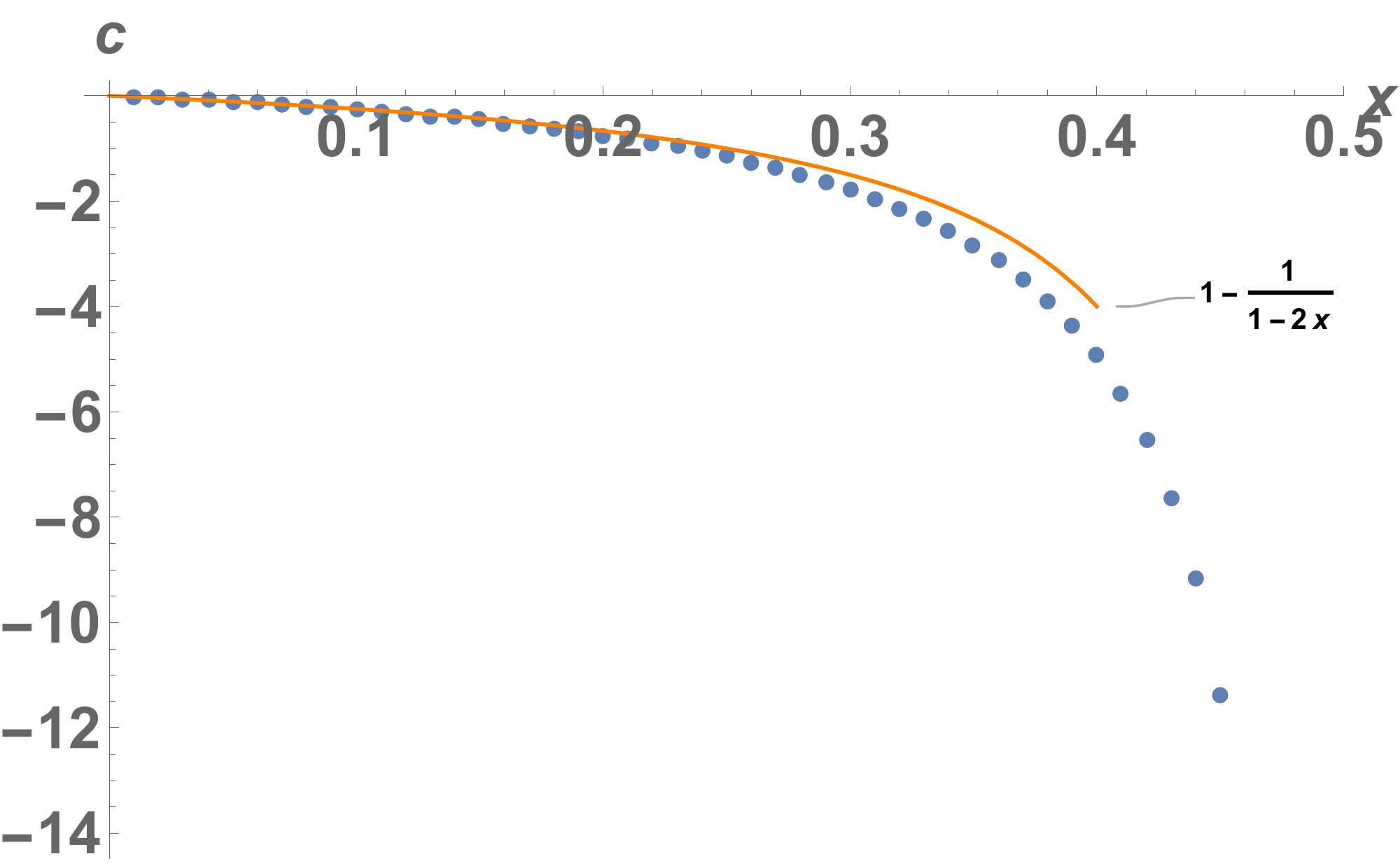}}
  \caption{The coefficient of $\dfrac{1}{N}$ in the two-loop corrections 
  as a function of temperature ($x=e^{-\frac{\omega}{k_B T}}$), for the case $b=f=1$. There is a divergence at the Hagedorn temperature.}
\label{cvsx}
\end{figure}

The trend of negative values of corrections breaks down at the Hagedorn temperature. There the correction is positive and seems to be independent of $N$. Above the Hagedorn point the uniform distribution is no longer the maximizing distribution. This abrupt change indicates the onset of the high-temperature regime, where one has to obtain a new formula for the vertex $V_{n_1,\cdots,n_p}$. In the low-temperature regime the finite $N$ corrections vanish as $\sim
\dfrac{c (x)}{N}$. One may extract $c (x)$ by computing $\displaystyle\lim_{N
\to \infty} N (\Lambda_3 + \Lambda_4)$. In Fig. [\ref{cvsx}] we
have shown the temperature dependence of this coefficient. $c (x)$ decreases with temperature and seems to diverge at the Hagedorn temperature. A crude curve-fitting exercise indicates that for smaller temperatures
\begin{align}
c (x) &\approx 1- \dfrac{1}{1-2x} & x &\sim 0
\end{align}
However, an analysis of log-log plot near the Hagedorn point shows that the dependence of $c(x)$ on $x$ is not a simple power law.

In the previous section, we deduced from Figs. [\ref{cubic_diag}],
[\ref{cubvsN}], [\ref{quartic_diag}] \& [\ref{qrtvsN}] that the
summands of $\Lambda_3$ and $\Lambda_4$ each were $\mathcal{O} (N^{-
2})$. Analyzing the cubic and quartic vertices we can see why this is so. For
simpler analysis, we shall keep $n_1$ and $n_2$ finite and make $N$ large.
When $N$ is large,
\begin{align}
  \tilde{s} \equiv s\mod N& = \begin{cases}
    s & s \geqslant 0\\
    N + s & s < 0
  \end{cases} & x^N &\approx 0
\end{align}
Using these simplifications, the cubic vertex becomes
\begin{dmath}
  V_{n_1, n_2, - n_1 - n_2} \approx - i \sqrt{N}  \left\{ n_1^2  \text{sgn}
  (n_1)  (1 - bx^{|n_1 |} + f (- x)^{|n_1 |}) + n_2^2  \text{sgn} (n_2) 
  \left( 1 - bx^{|n_2 |} + f (- x)^{|n_2 |} \right) \\
    - (n_1 + n_2)^2  \text{sgn} (n_1 + n_2)  (1 - bx^{|n_1 +
  n_2 |} + f (- x)^{|n_1 + n_2 |}) \right\} \\
    = - i \sqrt{N} \left\{ n_1^2  \text{sgn} (n_1) I_{| n_1 |}
  + n_2^2  \text{sgn} (n_2) I_{| n_2 |} - (n_1 + n_2)^2  \text{sgn} (n_1 +
  n_2) I_{| n_1 + n_2 |} \right\}
\label{Nvertex3}
\end{dmath}
while the quartic vertex looks like
\begin{dmath}
  V_{n_1, n_2, - n_1, - n_2} \approx \left\{ 2 | n_1 |^3  (1 - b x^{| n_1 |} +
  f (- x)^{| n_1 |}) + 2 | n_2 |^3  (1 - b x^{| n_2 |} + f (- x)^{| n_2 |})
  \\
    - | n_1 - n_2 |^3  (1 - b x^{| n_1 - n_2 |} + f (- x)^{|
  n_1 - n_2 |}) - | n_1 + n_2 |^3 (1 - b x^{| n_1 + n_2 |} + f (- x)^{| n_1 +
  n_2 |}) \right\} \\
    = 2 | n_1 |^3 I_{| n_1 |} + 2 | n_2 |^3 I_{| n_2 |} - | n_1
  - n_2 |^3 I_{| n_1 - n_2 |} - | n_1 + n_2 |^3 I_{| n_1 + n_2 |}
\label{Nvertex4}
\end{dmath}
From Eqs. [\ref{Nvertex2}, [\ref{Nvertex3}] and
[\ref{Nvertex4}] it becomes clear that the summands on the rhs of Eqs. [\ref{sum_cub}]
and [\ref{sum_qrt}] go as\footnote{The $n$ in the numerator simply indicates
the superficial power of mode number in the summand. It could e.g., represent
a factor like \unexpanded{$\dfrac{n_1^2}{n_2}$}.} $\sim \dfrac{n}{N^2}$. This confirms our
guess regarding the asymptotic $N$ dependence of each summand.

However, this also creates a new complication. Such a dependence on $n$ 
should lead to an $\mathcal{O} (N)$ divergence\footnote{This 
apparent $\mathcal{O} (N)$ divergence is not unique to two-loop case 
that is being considered here. A simple power counting indicates 
that it is present at higher-loop orders, too.} for each of $\Lambda_3$
 and $\Lambda_4$ after the double sum over mode numbers is 
performed. This is clearly against the numerical evidence that we have
at hand. From Eqs. [\ref{sum_cub}] and [\ref{sum_qrt}] we expect each
summand to go as $\sim \dfrac{1}{N^2 n}$ instead of the $\sim \dfrac{n}{N^2}$
that we see here. While normalization can be expected to remove the leading
divergent pieces (the ones that go as $\sim \dfrac{n}{N^2}$), it is not at all
clear how the subleading pieces that go as $\sim \dfrac{1}{N^2}$ get
canceled. For that one has to take a closer look at all the pieces in $\Lambda_3$ and $\Lambda_4$. 
The key lies in the $I_{|n|}$'s. They come with a $1$ and exponential 
convergent factors: $x^{| n |} = e^{-
\beta \omega | n |}$. All pieces whose numerators go as $\sim x^{| n_1 | + |
n_2 |}$ are absolutely convergent upon the double sum over $n_1$ 
and $n_2$. Only the $1$ in the $I_{|n|}$ could lead to divergences. For example, pieces that have this exponential convergence in only one of the
modes, i.e., pieces that go as $\sim x^{| n_1 |}$ or $\sim x^{| n_2 |}$, 
may be divergent (or convergent depending upon the power of the other 
mode). The worst fate is for the pieces whose numerators have no
convergence factor in either $n_1$ or $n_2$. However, one must keep in mind
that such pieces may ultimately get their divergences removed by proper
normalization.

\begin{table}[h]
  $\begin{array}{|l|}
    \hline
    \dfrac{n_1^3  \left( 2 \text{sgn} (n_1) - \text{sgn} (n_1 - n_2) -
    \text{sgn} (n_1 + n_2) \right)}{8 N^2 | n_1 | | n_2 | I_{| n_1 |} I_{| n_2
    |}}\\
    \\
    \dfrac{3 n_1^2 n_2  \left( \text{sgn} (n_1 - n_2) - \text{sgn} (n_1 + n_2)
    \right)}{8 N^2 | n_1 | | n_2 | I_{| n_1 |} I_{| n_2 |}}\\
    \\
    \dfrac{3 n_1 n_2^2  \left( - \text{sgn} (n_1 - n_2) - \text{sgn} (n_1 +
    n_2) \right)}{8 N^2 | n_1 | | n_2 | I_{| n_1 |} I_{| n_2 |}}\\
    \\
    \dfrac{n_2^3  \left( 2 \text{sgn} (n_2) + \text{sgn} (n_1 - n_2) -
    \text{sgn} (n_1 + n_2) \right)}{8 N^2 | n_1 | | n_2 | I_{| n_1 |} I_{| n_2
    |}}\\
    \hline
  \end{array}$
  \caption{The pieces that carry a superficial divergence in the summand of the quartic ``infinity'' diagram. Each individual term shows a divergence even after proper normalization. The combined expression, however, vanishes at large $N$.}
\label{qrtpiece}
\end{table}

All the divergent pieces in the quartic diagram are listed in the TAB
\ref{qrtpiece}. Each of the three pieces in the first line contains an
$\mathcal{O} (N)$ divergence on its own\footnote{This is even after
normalizing each piece properly.}. However, when all the terms on the first
line are taken together the superficial divergence is removed. A similar cancellation is exhibited by the terms on the fourth line as well. The first terms
on lines three and four combine to produce a vanishing contribution.
The second terms on those lines too behave in a similar way.

\begin{table*}[h]
  $\begin{array}{|c|}
    \hline
\begin{array}{ll}
  \dfrac{n_1^4  \left( \text{sgn} (n_1) - \text{sgn} (n_1 + n_2) I_{|n_1 + n_2
  |} \right)^2}{- 12 N^2 | n_1 | | n_2 | | n_1 + n_2 | I_{| n_1 |} I_{| n_2 |}
  I_{| n_1 + n_2 |}} & \dfrac{n_2^4  \left( \text{sgn} (n_2) - \text{sgn} (n_1
  + n_2) I_{|n_1 + n_2 |} \right)^2}{- 12 N^2 | n_1 | | n_2 | | n_1 + n_2 |
  I_{| n_1 |} I_{| n_2 |} I_{| n_1 + n_2 |}}\\
  & \\
  \dfrac{4 n_1^3 n_2  \left( I_{|n_1 + n_2 |}^2 - \text{sgn} (n_1)  \text{sgn}
  (n_1 + n_2) I_{|n_1 + n_2 |} \right)}{- 12 N^2 | n_1 | | n_2 | | n_1 + n_2 |
  I_{| n_1 |} I_{| n_2 |} I_{| n_1 + n_2 |}} & \dfrac{4 n_1 n_2^3  \left(
  I_{|n_1 + n_2 |}^2 - \text{sgn} (n_2)  \text{sgn} (n_1 + n_2) I_{|n_1 + n_2
  |} \right)}{- 12 N^2 | n_1 | | n_2 | | n_1 + n_2 | I_{| n_1 |} I_{| n_2 |}
  I_{| n_1 + n_2 |}}\\
\end{array}\\
\\
    \dfrac{2 n_1^2 n_2^2  \left( 3 I_{|n_1 + n_2 |}^2 - \text{sgn} (n_1) 
    \text{sgn} (n_1 + n_2) I_{|n_1 + n_2 |} - \text{sgn} (n_2)  \text{sgn}
    (n_1 + n_2) I_{|n_1 + n_2 |} + \text{sgn} (n_1) \text{sgn} (n_2)
    \right)}{- 12 N^2 | n_1 | | n_2 | | n_1 + n_2 | I_{| n_1 |} I_{| n_2 |}
    I_{| n_1 + n_2 |}}\\
    \hline
  \end{array}$
  \caption{The pieces that carry a superficial divergence in the summand of the double-cubic
  theta diagram. Each individual term shows a divergence even after proper normalization. The combined expression, however, is vanishing for large $N$.}
\label{cubpiece}
\end{table*}

In TAB. [\ref{cubpiece}) all the divergent pieces of the double-cubic diagram have been listed.
However, unlike the quartic case, there are many more pieces. One can
check that each of the five expressions gives rise to a $\mathcal{O} (N)$
divergence, even after proper normalization. It is only when all five are combined that these divergences finally get removed.

\section{Conclusion}
In this paper we discussed an algorithm for calculating $V_{n_1, \cdots,
n_p}$ for finite $N$. This algorithm is pivoted on the fact that
$\displaystyle\sum_\theta \dfrac{d}{d \theta} e^{t + i \theta} = i \dfrac{d}{d
t}  \displaystyle\sum_\theta e^{t + i \theta}$. This enables one to first sum
over different values of $\theta$ and then take derivatives with respect to
a different variable. In {\cite{curtright_color_2017}}, one did not need to do
this as the sum was approximated by an integral, which was subsequently solved
using integration by parts. The main limitation of our algorithm is that it is valid only when the uniform distribution is the global maxima of $L$. Above the Hagedorn temperature the maximizing distribution starts depending on the temperature. It would be an interesting exercise to derive a compact expression for $V_{n_1,\cdots,n_p}$ for  $x>\dfrac{1}{b+f}$.

An analysis of the steepest descent method demonstrates that the integral of the Haar measure for $S U (N)$ is not approximated
well by the Gaussian fluctuations about its maxima. There are nonvanishing corrections due to two-loop diagrams. We obtained numerical
evidence for a small $\mathcal{O} (N)$ contribution from even the higher-loop corrections. This means that any calculation of $\log (Z)$ may
also have $\mathcal{O} (N)$ remnants if one stops at the Gaussian fluctuations.
The (bare) $3$-vertex and the $4$-vertex can potentially
contribute to $\mathcal{O} (N)$ terms. Obtaining a closed-form expression for the large $N$ dependence for the two-loop corrections to the Haar measure will be an interesting endeavor for the future.

In order to do a detailed study, we obtained general expressions for the bare cubic and quartic vertices for finite $N$. We
then computed corrections to the log of the partition function due to
two-loop diagrams. From the expressions of the summands of the two-loop
diagrams it is not at all obvious whether the mode sums would vanish as $N$
becomes large. We proceeded to check this numerically and found that the contribution from
the double-cubic and quartic terms are $\mathcal{O} (N^{- 1})$ and hence
indeed vanish as $N \rightarrow \infty$. A study of each diagram showed that
every superficially divergent piece in those diagrams is canceled by another
superficially divergent piece. This made each of $\Lambda_3$ and $\Lambda_4$ negligible
compared to the Gaussian approximation. The diverging pieces in $\Lambda_4$ 
that cancel each other were identified in this paper. It would be 
instructive to inspect and repeat that analysis for similar pieces in 
$\Lambda_3$. The total two-loop correction is negative below 
the Hagedorn point, which is consistent with the expectation for 
finite $N$ partition functions. The coefficient of the $\dfrac{1}{N}$ 
corrections shows a monotonic decrease with temperature, 
with an indication of a divergence at the Hagedorn point. The analytic 
dependence of this coefficient was not obtained in this paper. It 
will be an interesting exercise to obtain this dependence from 
analytic, closed-form expressions for the two-loop, finite $N$ corrections.

\section*{Acknowledgments}
I would like to thank Charles Thorn for his insights and guidance in 
this project. This work was supported in part by the Department of 
Energy under Grant No. DE-SC0010296.

%

\end{document}